%% file: gravity_220805.tex
\documentclass[multphys,vecphys]{svmult}

\usepackage{amssymb} 
\usepackage{makeidx}         
\usepackage{graphicx}        
\usepackage{multicol}        
\usepackage[bottom]{footmisc}

\makeindex             

\begin{document}

\title*{GRAVITY: The AO-Assisted, Two-Object Beam-Combiner Instrument
for the VLTI}
\titlerunning{GRAVITY: The AO-Assisted, Two-Object Beam-Combiner Instrument}
\author{F.~Eisenhauer\inst{1} \and G.~Perrin\inst{2} \and
S.~Rabien\inst{1} \and A.~Eckart\inst{3} \and P.~L\'ena\inst{2} \and
R.~Genzel\inst{{1,4}} \and R.~Abuter\inst{1} \and T.~Paumard\inst{1}
\and W.~Brandner\inst{5}}
\authorrunning{Eisenhauer et al.}
\institute{Max-Planck-Institut f\"ur extraterrestrische Physik (MPE),
Garching, Germany \and Observatoire de Paris -- site de Meudon, France
\and 1. Physikalisches Institut der Universit\"at K\"oln, Germany \and
Department of Physics, University of California, Berkeley, USA \and
Max-Planck-Institut f\"ur Astronomie (MPIA), Heidelberg, Germany}
\maketitle

\abstract{We present the proposal for the infrared adaptive optics
(AO) assisted, two-object, high-throughput, multiple-beam-combiner
GRAVITY for the VLTI. This instrument will be optimized for
phase-referenced interferometric imaging and narrow-angle astrometry
of faint, red objects. Following the scientific drivers, we analyze
the VLTI infrastructure, and subsequently derive the requirements and
concept for the optimum instrument. The analysis can be summarized
with the need for highest sensitivity, phase referenced imaging and
astrometry of two objects in the VLTI beam, and infrared
wavefront-sensing. Consequently our proposed instrument allows the
observations of faint, red objects with its internal infrared
wavefront sensor, pushes the optical throughput by restricting
observations to K-band at low and medium spectral resolution, and is
fully enclosed in a cryostat for optimum background suppression and
stability. Our instrument will thus increase the sensitivity of the
VLTI significantly beyond the present capabilities. With its two
fibers per telescope beam, GRAVITY will not only allow the
simultaneous observations of two objects, but will also push the
astrometric accuracy for UTs to 10 $\mu$as, and provide simultaneous
astrometry for up to six baselines. }


\section{Introduction}
\label{introduction}
GRAVITY is a general purpose instrument for the VLTI (Glindemann et
al. 2000) with two main capabilities: phase-referenced imaging, and
narrow-angle astrometry.  A non-exhaustive list of possible
applications is listed in section \ref{scientific justification}. The
design of the instrument is driven by a set of specific science cases,
that also set its name (\emph{General Relativity Analysis via
V{\scriptsize LT} InTerferometrY}, or GRAVITY), namely to explore the
so far untested regime of strong gravity encountered at a few hundred
down to a few Schwarzschild radii ($R_\mathrm{S}$) from supermassive
objects.


\section{Scientific Justification}
\label{scientific justification}
\subsection{The Center of the Milky Way}
\label{galactic center}
The Galactic Center is the closest galactic nucleus, and the best
candidate for a supermassive black hole (BH). However, the final proof
that Sgr~A* is smaller than its event horizon is still pending. Its
Schwarzschild radius is $R_\mathrm{S}\simeq9$~$\mu$as. With GRAVITY we
aim at probing the space-time around this object down to a few
$R_\mathrm{S}$. This goal is detailed in the contribution to this
workshop by Paumard et al. (2005a) and summarized
hereby. Unfortunately, the adaptive optics (AO) correction with a
visible wavefront-sensor is only modest (Strehl $\approx$ 10\% in
K-band), because no bright visible guide star is nearby. In contrast,
wavefront-sensing at infrared (IR) wavelengths allows observations
close to the diffraction limit at 2 $\mu$m (Strehl $\approx$
50\%). Another complication is that the emission from the central 100
mas around Sgr~A* is quite low ($m_\mathrm{K}\gtrsim17.5$), and that
Sgr~A* lies in a crowded region (a dozen of stars with
$m_\mathrm{K}\simeq15$ within a radius of $0.5''$, a few stars with
$m_\mathrm{K} \simeq $ 10--11 at $\simeq1''$).  To overcome all these
difficulties, it is necessary to have a photon-efficient instrument
and IR wavefront-sensing, i.e. GRAVITY. On the other hand, the high
density of stars can be turned into an advantage: the $2''$ field of
view (FoV) of a single VLTI beam will contain a sufficiently bright
star for phase referencing, which will allow us to achieve an
astrometric accuracy of approx. $10$~$\mu$as in only $1$--$2$~min
integration time.

{\bf Probing Space-Time around the Supermassive Black Hole}: Sgr~A* is
known for exhibiting so-called flares a few times a day. These events
last for about $1$~h, and their light-curves show significant
variations on a typical timescale of $17$~min in the IR. From the
typical raise time of the substructures in the light-curve, they have
to come from regions smaller than about $10$ light-minutes, or
$\simeq17$~$R_\mathrm{S}\simeq150$~$\mu$as.  The emitting region
cannot remain static in the potential well of the BH, but its velocity
will be comparable to the Keplerian circular velocity
$v\simeq(r/R_\mathrm{S})^{-1/2}\times10\:\mu\mathrm{as}\:
\mathrm{min}^{-1}\gtrsim1\mu\mathrm{as}\:\mathrm{min}^{-1}$.
Therefore measuring the 2D astrometry of flares with $10$~$\mu$as
accuracy and a time resolution of a few minutes will allow not only
determining the location of the flares with respect to the BH, but
also their proper motion. This will allow us to understand their
nature, and provide a probe in the potential well of Sgr~A* at a few
to $100$~$R_\mathrm{S}$. If the flares come from very close to the BH,
the $17$~min periodicity may be caused by the beaming of the
radiations emitted by particles on the last stable orbit of a Kerr BH
of spin parameter $0.52$ (Genzel et al. 2003).  If this is the case,
then $10$~$\mu$as-accuracy, minute-sampling astrometry will allow us
to trace the lensed image of the last stable orbit, and therefore to
probe the space-time down to the photon-sphere of the BH (Broderick \&
Loeb 2005, Paumard et al. 2005b).  This experiment absolutely requires
the simultaneous, multi-baseline, narrow-angle astrometry capability
of GRAVITY.

{\bf Probing Stellar Dynamics in the Regime of General Relativity}:
The current best estimates of the mass of the central BH and distance
to the Galactic Center are obtained through orbit-fitting of stars in
the central arcsecond of the Galaxy, the so-called S-stars. But
stellar counts predict that a few faint stars ($17.5\lesssim
m_\mathrm{K}\lesssim19.5$) should reside even within the central
$100$~mas of the Galaxy. These stars have orbital periods of order one
year, periapses of order $1\:\mathrm{mas}\simeq100$~$R_\mathrm{S}$,
and travel at relativistic velocities during their periapse passages.
The repeated interferometric imaging of these stars allows to test
relativistic effects, in particular the prograde periastron shift. In
addition, it will give the mass enclosed within
$\simeq100$~$R_\mathrm{S}$, while the S-stars measure the mass
enclosed within $\simeq1000$~$R_\mathrm{S}$: comparing the two numbers
will give a measurement of the mass of the stellar cusp. Note that the
observations for this project also provide the monitoring for flares
to perform the above ``flare'' experiment. The main advantage of
GRAVITY over competing instruments is its IR wavefront sensor, and its
optimized K-band throughput.

{\bf Astrometry of S-Stars}: The astrometric capability of GRAVITY
will also allow deriving very accurate orbits for the more distant
S-stars. Specifically when they pass periapse, we will get a
significant improvement on the determination of the position, mass,
and distance of Sgr~A*. These improvements are indeed required to
further constrain the modelling in the two aforementioned experiments.
On a longer time-scale (decade), the astrometry of the S-stars will
also allow finding the General Relativistic effects, and probing for
the extended mass component at the scale of one arcsecond. Compared to
the general PRIMA astrometric facility, where the light from the phase
reference star and the S-stars is unecessarily seperated at the UT
Coude focus, GRAVITY will profit from its narrow-angle astometric mode
with phase referencing within the 2'' single interferometric beam.

\subsection{Intermediate Mass Black Holes}
\label{imbh}
There is compelling evidence for a very compact cluster (GCIRS~13E)
that may contain an intermediate mass black hole (IMBH, Maillard et
al. 2004).  Interferometric imaging of the cluster will provide
reliable proper motions for the core stars, and hence dynamically test
the IMBH hypothesis. Another place to look for IMBHs is globular
clusters. The current main limitation on the ability to measure the
mass of these putative dark objects is the low number of bright stars
in the core of the Globular Clusters suitable for radial velocity
measurements (Baumgardt et al. 2005).  GRAVITY will improve the
situation by allowing high accuracy proper motion measurements.  At 8
kpc, $1$~km/s $\simeq26$~$\mu$as~yr$^{-1}$, meaning that this accuracy
will be reached in $1$~yr with narrow-angle astrometry or a few years
through imaging.  In addition, the acceleration of a star orbiting a
$1000$~$M_\odot$ BH at $4$~mpc ($100$~mas at $8$~kpc) is
$\simeq7$~$\mu$as~yr$^{-2}$, so that for such stars the acceleration
would be detected after a few years in astrometric mode.  The same
methods can also be applied to search IMBHs in young Galactic
starburst clusters (e.g. Arches). Like the experiment on the GC S-star
orbits, this science case profits tremendously from the unique
narrow-angle astrometry capability of GRAVITY.

\subsection{Stellar Orbits around Extragalactic Supermassive Black Holes}
\label{extragalactic orbits}
The nucleus of M31 is made of a blue disk of $200$~Myr old stars
orbiting a $\simeq1.4\times10^8\:M_\odot$ BH (Bender et al. 2005). The
half-power radius of this disk is $\simeq60\: \mathrm{mas}=0.2\:
\mathrm{pc}$, and it should contain of order $10$ bright red evolved
stars, that would be resolvable by GRAVITY. Their proper motion, of
order $\simeq1.7\times10^{3}\:\mathrm{km/s}
\simeq0.5\:\mathrm{mas~yr}^{-1}$ at $0.76$~Mpc, will be measurable
within only a few years by VLTI imaging. The same proper motion
measurements will also be possible in other nearby galaxies. Compared
to other instruments with strong spectroscopic capabilities, GRAVITY
will be superior for this project because of its optimisation for
single band, low-spectral resolution imaging.

\subsection{Active Galactic Nuclei}
\label{agn}
Many active galactic nuclei (AGN) are deeply dust-embedded, and faint
in the optical. For these object the IR wavefront-sensing capability
of GRAVITY is mandatory. High spectral resolution is not necessary for
many projects, because the spectral features of AGNs are comparably
broad. More important is the high sensitivity of GRAVITY, a
consequence of the optimization for single-band operation at low
spectral-resolution. In the standard model of AGNs, the centers of
galaxies host supermassive BHs with masses in the range $10^6 - 10^9
M_\odot$. The large luminosity of the core is of gravitational origin,
with large amounts of material orbiting in an accretion disk feeding
the central BH. The disk and BH are enshrouded in a dust torus beyond
the condensation radius for dust. The central part of the AGN contains
the Broad Line Region (BLR), a compact region in which the gas
velocity can be as large as a few 1000 km/s. Typical scales in Seyfert
galaxies for these basic components of the AGN theory are as follows:
BH Schwarzschild radius $R_\mathrm{S} = 10^{-5} \left(
\frac{M_{BH}}{10^8 M_\odot} \right)$ pc, jets up to several kpc, outer
edge of accretion disk typically 1000 $R_\mathrm{S}$, size of BLR
about 0.003--0.3 pc, dust torus inner radius $\sim$1 pc. The accretion
disk is of the order of 0.2~mas (0.01~pc) and is beyond reach for
VLTI. Up to now the BLR sizes are indirectly derived from
reverberation mapping measurements (e.g. Kaspi et al.  2000) in the
blue part of the visible spectrum. The sizes of dust tori are guessed
from the distance for which graphite and silicate grains can condense
(e.g. Barvainis 1987) or by modeling the IR part of the SED, in both
cases indirect methods.  The size of the torus of NGC~1068 has only
recently been directly measured with MIDI at VLTI (Jaffe et al.
2004). Near-infrared (NIR) observations are sensitive to both the dust
torus, whose emission in that wavelength range at least for Seyfert 2
remains large, and the more compact central source. Spectroscopically
resolved observations can allow to disentangle between regions with
spectral features such as the BLR and continuum sources such as the
dusty torus or the central engine. In the BLR, lines are as broad as a
few 1000~km/s, and a medium spectral resolution of 750 makes the BLR
identification easy. The dust torus will a priori be completely
resolved for sources closer than 100~Mpc. The largest BLRs are within
reach at NIR wavelengths with angular scales of 2~mas for sources
closer than 100~Mpc. The Br$\gamma$ and Pa$\alpha$ emission lines
can be used to detect the BLR in the K-band.  Br$\gamma$ might also
trace any shock in the circum-nuclear environment. This may provide an
efficient tool to detect the base of the jets where they are launched
into orbiting material. The coupling of imaging and spectroscopy as
proposed by Woillez et al. 2005 will allow to combine imaging and
reverberation mapping to obtain, through tomography, the 3D structure
of the BLR. A by-product is the direct measurement of the distance of
AGNs by comparing the angular and linear sizes of the BLR, a
measurement of cosmological significance (Elvis \& Karovska 2002). A
preliminary source selection based on distance, brightness and nearby
reference source availability points towards four main sources
(NGC~1068, Circinus, NGC~3783, NGC~3758) whose core magnitudes are
brighter than K=13. All four sources have bright reference stars (K
brighter than 9.4) less than 2~arcmin away.

\subsection{Stars and Starformation}
\label{stars}
{\bf Masses of the Most Massive Stars}: There still exists a
discrepancy by up to a factor of 2 in the mass estimates for the most
massive main-sequence stars. Comparison of spectra with atmospheric
models yields upper mass limits in the range of 60\,M$_\odot$, whereas
evolutionary tracks and observed luminosity suggest a mass of up to
120\,M$_\odot$ for stars of spectral type O2V and O3V.  Clearly,
dynamical mass estimates for early O-type main-sequence stars are
required. Luckily, quite a number of spectroscopic binary O-stars are
known in the cores of Galactic starburst clusters like Arches,
Quintuplet of NGC 3603, or extragalactic starbursts like 30\,Dor.
With a nominal resolution of 4\,mas at a wavelength of 2\,$\mu$m,
Gravity could resolve some of the longer period spectroscopic
binaries, and monitor the astrometric motion of the photocenter for
the shorter period, closer binaries. Astrometric orbits for these
deeply embedded binary stars will hence directly yield dynamical mass
estimates. The unique narrow-angle astrometry mode of GRAVITY is ideal
for these dynamical studies in crowded regions. Several objects
(e.g. Arches, Quintuplet) can not be observed without the IR wavefront
sensing provided by GRAVITY.

{\bf Circumstellar Disks and Jets around Young Stars}: Circumstellar
disks and outflows are closely linked to the star formation
process. The presence of a circumstellar disk is also the
pre-requisite for the formation of planetary systems. The
gravitational interaction between a planet and a disk should manifest
itself in the occurrence of spiral structures, wakes and gaps. Such
structures have indeed already been observed in a number of cases
(e.g. GG Tau, Formalhaut, etc.). The relative faintness of a young
giant planet compared to the high-surface brightness of a typical
circumstellar disk thus far prevented its direct detection in
diffraction limited observations with 8m class telescopes in the NIR.
GRAVITY's 4\,mas resolution (compared to 60\,mas resolution for one
UT) drastically improves the contrast between a disk and its embedded
planet by a factor of (60/4)$^2$ = 225. Hence it should be possible to
probe for young giant planets which are almost 6\,mag fainter than
what is currently achievable. While the ubiquity of jets in star
formating regions has been well established, the physics behind the
formation of jets, and in particular the launching mechanism, is still
poorly understood. Models (e.g. Turner et al. 1999) suggest that first
angular momentum gets transferred along horizontal magnetic field
lines from the disk to the central material, which then gets
accelerated along a vertical pressure gradient, ultimately forming a
collimated jet. The important processes seem to take place within less
than 0.5 AU from the star, which at typical distances to the nearest
star forming regions of 150 pc translates into an angular size of less
than 30 mas. At 4\,mas resolution, GRAVITY will be able to resolve the
central jet formation engine around young, nearby stars. Furthermore,
at a distance of 150 pc, an astrometric precision of 10 $\mu$as
corresponds to a transversal velocity of $\approx$ 60\,km/s. Hence
high-velocity outflows, and the formation and evolution of jets from
T\,Tauri stars with typical velocities of 150\,km/s can be resolved
and traced in real-time. These observations will put tight constraints
on jet formation models and the role of magnetic fields.

\subsection{Planets and Multiple Systems}
\label{planets}
{\bf Substellar Objects in Multiple System - Dynamical Masses and
Calibration of Theoretical Models}: Recent claims on the direct
detection of planetary mass companions to young brown dwarfs (2MASSW
J1207334-393254, Chauvin et al. 2005) and low-mass stars (GQ Lup,
Neuh\"auser et al. 2005) are based on evolutionary models and model
atmospheres, which have not yet been accurately calibrated to
observations.  Dynamical mass estimates for ultra-cool dwarfs and
brown dwarfs have thus only been derived for a handful of objects (LHS
1070, GJ 569Bab, 2MASS J0746+2000, AB Dor C). In general, the observed
masses for substellar objects with ages larger than a few 100 Myr seem
to be in good agreement with theoretical models. AB Dor C, however,
the close companion to the K7 Zero-Age-Main Sequence star AB Dor, is
among the youngest very-low-mass objects for which a precise mass
estimate has been obtained (Close et al. 2005). Quite surprisingly, AB
Dor C turned out to be about twice as massive as would have been
expected from theoretical models for an age of $\simeq$ 40\,Myr. If
this result should hold true, it has profound implications on the
substellar mass functions of young stellar systems, and would also put
the existence of the so-called ``cluster planets'' into doubt (71 sig
Ori, e.g., would then well be in the brown dwarf and not the planetary
mass regime).  Clearly, more dynamical (astrometric) mass estimates
for young, very-low mass and substellar objects are required in order
to calibrate the theoretical tracks and model atmospheres. GRAVITY
could probe many more multiple systems like AB Dor, which have at
least one very-low mass or substellar component.  GRAVITY will yield
astrometric orbits, and derive the individual component
masses. Furthermore, GRAVITY allows one to probe the substellar
companions themselves for binarity. In the case of AB Dor C, the only
way to reconcile present day evolutionary tracks with the dynamical
mass would be if AB Dor C would be a close binary, consisting of two
40 M$_{\rm Jup}$ brown dwarfs.  Finally, GRAVITY could even resolve
the motion of details (such as atmospheric clouds) on the planet
surface, provided this detail dominates the photo-center of the
planet: $1$ Jupiter diameter would be $10$~$\mu$as at $100$~pc.

{\bf Planets in binary systems}: The Sun's wobbling due to Jupiter is
of order $1\:R_\odot$, or $1$~mas at $10$~pc, or $10\:\mu$as at
$1$~kpc. If the Sun were a double star (true binary or in projection),
so that the companion could be used as a phase reference, GRAVITY
would be able to detect Jupiter from $1$~kpc away in one Jupiter
period ($12$~yr).  The wobble of a Sun due to a hot Jupiter could be
determined at distances up to $\simeq10$~pc in a few days. GRAVITY
offers the possibility to search for exoplanets at kpc distances even
with the ATs thanks to its optimised thoughput. Finally, if the planet
does not have the same color as the star, the photo-center should not
be the same at both ends of the K band. This difference would give a
handle on the linear scale of the system.

\subsection{Microlensing}
\label{microlensing}
Gravitational microlensing events occur when a point-like source and a
massive object (lens) are almost aligned on the line of sight.  This
causes three effects: the luminosity of the source appears enhanced,
its photo-center is slightly displaced, and a secondary image appears
on the other side of the lens.  These events are interesting to probe
the mass spectrum of compact objects in the Galaxy, but the
light-curves currently provided by the dedicated surveys are not
enough to determine the mass: 2D spatial information (either 2D track
of the photo-center or separation between the two images) is necessary
to disentangle it from the other event parameters.  GRAVITY would be
very efficient in following events reported by photometric surveys,
because its high sensitivity would put a lot of events in reach of the
ATs, and most of them in reach of the UTs. The usual events have a
simple geometry and are often long ($>10$ days).  However, when the
lens is not simple but has a companion (similar dark object or
planet), caustic crossings can happen, characterized by multiple
images and complex wobbling of the photo-center. Both astrometry and
imaging can give access to the mass of the companion. The second of
two crossings (which usually go by pair) can be predicted a few days
in advance, but it lasts only a few hours. Therefore GRAVITY's ability
to measure high precision phases and visibilities in minutes is highly
valuable.


\section{Top Level Requirements in the Context of the VLTI Facility}
\label{requirements}
In this section we detail the top-level requirements from the
scientific justification (section \ref{scientific justification}), and
outline that the present VLTI with its first generation of instruments
and the upcoming PRIMA facility (Delplancke et al. 2000) does not
fulfill these functional and performance needs (see Wilhelm et
al. (2002) for the functional description of the VLTI, and Delplancke
(2004b) for a PRIMA executive summary).

\subsection{Functional Requirements}
\label{function}
\begin{itemize}
\item IR wavefront sensing for observations of red objects: The four
  UTs are equipped with the MACAO adaptive optics (Arsenault et
  al. 2003), which sense the wavefront at visible wavelengths. No IR
  wavefront-sensor is available at the UTs. The ATs have no high order
  AO at all.
\item Simultaneous astrometry for multiple baselines: The present
  PRIMA facility offers two star-separators (STSs) for UTs (plus two
  for ATs), two fringe sensor units (FSUs), and two differential delay
  lines (DDLs). In the near future (2006), at least three UTs will be
  equipped with STSs, but the astrometric mode of PRIMA will still be
  limited to a single baseline per observation. The future upgrade
  of PRIMA for the second generation instruments with four FSUs and
  four DDLs potentially allows simultaneous multiple baseline
  astrometry, but details are not settled.
\item Narrow-angle astrometry for distances less than two arcseconds:
  The PRIMA star-separators are optimized for angular separations
  between the primary star and the secondary object larger than
  2$''$. The star-separators are not optimum for narrow-angle
  astrometry. The 0--2$''$ range can be reached as well, but only
  with the risk of cross-talk between the two channels (Delplancke
  2004b).
\item Phase-referenced imaging interferometry using multiple
  telescopes: The present PRIMA facility will allow phase-referenced
  imaging for three telescopes. The $2^{nd}$ generation PRIMA will
  then co-phase four telescopes, necessary for phase-referenced imaging
  on six baselines.
\end{itemize}

\subsection{Performance Requirements}
\label{performance}
\begin{itemize}
\item AO corrections achieving near-diffraction limited
  performance (Strehl $\simeq$ 50\%) for stars with a K-band
  magnitude $m_\mathrm{K}$ $\ge$ 10.
\item Narrow-angle astrometry with $\simeq$ 10 $\mu$as accuracy
  in five minutes observing time: While the astrometric accuracy with
  the ATs will be $\simeq$ 10 $\mu$as, the more complex and longer
  non-common optical light path for UTs reduces the expected accuracy
  to $\simeq$ 100 $\mu$as (Delplancke 2004b). The astrometric
  accuracy of PRIMA using the UTs is thus expected to fail our top
  level requirements by approximately a factor 10.  The random
  atmospheric differential OPD residuals do not limit significantly
  the accuracy for narrow-angle astrometry. Even for observations as
  short as 3 minutes, the typical average atmospheric OPD residual is
  less than 5 nm or 10 $\mu$as.
\item Phase-referenced imaging interferometry with a point source
  limiting K-band magnitude of $m_\mathrm{K}$ $\ge$ 19 in one hour
  observing time: The limiting magnitude of the first generation AMBER
  instrument (Petrov et al. 2000) in K-band without external fringe
  tracking is presently $m_\mathrm{K}$ $\simeq$ 7 in 25 ms
  (Rantakyr\"o 2005). Even with external fringe-fracking with PRIMA,
  the present VLTI and its $1^{st}$ generation instruments will miss
  our sensitivity requirement for phase-referenced imaging
  ($m_\mathrm{K}$ $\simeq$ 19 in one hour) by approximately 2
  magnitudes. The main reason for the limited performance is the small
  optical throughput of $\simeq$ 2\% (Rantakyr\"o 2005) of AMBER,
  resulting from the trade-off necessary to provide a multi-purpose,
  multi-wavelength instrument for reasonable cost.  In contrast, a
  simple throughput-optimized broad-band beam-combiner like the PRIMA
  FSU is expected to go down to 19$^{th}$ mag (Delplancke 2004b).
\end{itemize}

We conclude that only a dedicated instrument will allow the
astronomical and physical key-experiments proposed in section
\ref{scientific justification}. In short, we need an IR-AO assisted,
two-object beam-combiner instrument, optimized for highest optical
throughput, most accurate multi-baseline narrow-angle astrometry, and
phase-referenced imaging.


\section{GRAVITY: Concept and Observing Modes}
\label{concept and modes}

\subsection{An IR AO Assisted, Two-Object Beam-Combiner Instrument
for the VLTI}
\label{concept}
The concept of GRAVITY is directly derived from the top-level
requirements. The instrument will be installed in the VLTI
laboratory. It will use the 18 mm diameter input-beams of the four
telescopes as provided by the beam compressor unit (Koehler \& Gitton
2002), and physical access to the exit pupils of the VLTI. The whole
instrument will be enclosed in a cryostat and evacuated. This will not
only dramatically improve the stability and cleanliness of the
instrument, but will also allow for optimum baffling and suppression
of thermal background. In our concept we foresee the option to
separate the AO module for use with other VLTI instruments. In its
baseline design, GRAVITY will only have IR wavefront-sensors, and
command the deformable mirrors (DMs) of the MACAOs. Alternatively we
consider a full AO system including DMs in the VLTI laboratory. The
second major sub-system of GRAVITY is the field-selector unit, and is
best described as a highly compact PRIMA facility working on a single
interferometric beam.  This unit picks two objects from the 2$''$ FoV
and couples the light to single-mode fibers for beam cleaning. The
relative optical path difference (OPD) of the two objects will be
compensated by stretching the fibers. An internal metrology system
will measure the OPD of the two beams. In contrast to the PRIMA
facility, however, the OPD control of the GRAVITY field-selector is
significantly easier, because in GRAVITY the accuracy and stability of
5 nm (corresponding to an astrometric accuracy of 10 $\mu$as) is only
required within a single interferometric beam, and not between the two
beams of the PRIMA facility. The feasibility of 10 $\mu$as narrow
angle astrometry with fringe-tracking on a single beam has already
been demonstrated at the Palomar Testbed Interferometer (Lane \&
Muterspaugh 2004). The light from the eight fibers (2 objects from 4
telescopes) is then fed to the GRAVITY beam-combiner, the third major
sub-system of the instrument. Three options are presently investigated
for optimum throughput, stability, cost, and complexity: Integrated
optics, fiber X coupler, and bulk-optics beam-combiners. The last
major sub-system in GRAVITY is the camera unit. Optimized for highest
possible throughput, we will only implement transmissive prisms and
grisms for low (R $\simeq$ 20) and medium (R $\gtrsim$ 500) spectral
resolution. Various detector options are presently investigated, with
focus on large pixels, high quantum-efficiency, and low noise.

\subsection{Basic Observing Modes}
\label{modes}

GRAVITY will provide two basic observing modes: {\bf Simultaneously
phase-referenced imaging of up to two objects}, and {\bf narrow-angle
astrometry within the 2$''$ FoV of the VLTI}.  GRAVITY will provide
{\bf internal fringe-tracking} for reference stars within the 2'' FoV,
but can also be operated using external fringe-tracking with PRIMA for
more distant reference stars.  GRAVITYs advantage over the PRIMA
astrometric mode is the ability of simultaneously measuring multiple
baselines, and the largely increased accuracy for UT operation because
of the restriction to small distances and operation in a single
interferometric beam. Because the instrument will measure the
visibility, phases and differential OPD of the two objects for {\bf
all six baselines}, GRAVITY will also track on significant fainter
sources than PRIMA, which operates only on the minimum subset of
baselines.  GRAVITYs advantage over competitive instruments for
phase-referenced imaging is the sensitivity-optimized concept. We do
not compromise for multi-mode, multi-wavelength and high spectral
resolution. GRAVITY also has the multiplex advantage from two objects
per FoV. This implies the possibility to use one of the two objects
for simultaneous visibility calibration. And GRAVITY will have the IR
wavefront-sensors for dust-obscured objects.

\section{Key Components of GRAVITY}
\label{key components}
In this section we briefly outline the possible options for the
various key-components of GRAVITY. The technology for all sub-systems
is well advanced, and (semi-) commercial devices are available. There
is no technological show-stopper.

\subsection{Infrared Wavefront-Sensor and Deformable Mirrors}
\label{ao}
IR wavefront sensors are already installed in various astronomical
facilities (e.g. NAOS, Gendron et al. 2003), typically as
Shack-Hartmann sensors using lenslet-arrays. Also the concept of
pyramid wavefront-sensors is promising for GRAVITY. A first IR device
has recently been installed at the Calar Alto observatory (Costa et
al. 2004). Curvature wavefront sensors could be advantageous, because
they could be matched optimally to the MACAO DM geometry (GRAVITY
option without additional DMs). The default detector for the wavefront
sensor would be the Rockwell HAWAII detector with a read-noise of
typically 10 electrons (Finger et al. 2002). The recent developments
for large-pixel, low-noise detectors (e.g. Rockwell CALICO prototype,
Finger et al. 2004) for IR wavefront-sensors may allow to increase the
sensitivity by a factor of two (0.75 mag). Optionally GRAVITY will be
equipped with its own DMs (not commanding the MACAO DMs). In this case
micro-machined DMs will be used directly in the 18 mm VLTI
beam. Potential technologies include magnetic DMs (e.g. from
LAOG/LETI, Cougat et al. 2001), or piezo-electrically driven mirrors
(e.g. from OKO Technologies, Dayton et al. 2000).

\subsection{Field-Selector, Pathlength-Compensation, and Metrology}
\label{field selector}
In order to not move the fibers during observation, we foresee a
K-mirror for de-rotation, a tip-tilt for laterally moving the field,
and a device to adjust the projected separation of the two fibers. A
fiber streching unit will introduce a differential OPD to one of the
objects. All motions may be implemented for cryogenic operation, for
example using stepper motors (e.g. Berger Lahr) and inductive
resolvers (e.g. LTN Servotechnik) for the K-mirror, piezo stacks with
capacity sensors for the tip-tilt mirror and OPD control (e.g. PI
Ceramic), and piezo-electric translation stages (e.g. Attocube
Systems) for fiber coupling. The standard solution for measuring the
differential OPD is a dual wavelength laser metrology (e.g. PRIMET,
Leveque et al. 2003). The metrology beams will be launched directly at
the beam-combiner. Another option for the metrology is to launch the
laser at the center of the pupil at wavelengths just above and below
the operational wavelength range, and track its phase with the science
beam combiner. For that we would make use of the flexible read-out
technology of IR detectors, which allow kHz sampling for the
metrology pixel, while integrating on the science object.

\subsection{Optimization for Faint Objects}

The strategy for minimizing light loss is two-fold: reduce the number
of optical elements and optimize their losses: first we use reflective
optics instead of transmissive optics, and combine several functions
in one unit (e.g. field-selector is also the fiber-coupler
optics). Second, we restrict our wavelength range to K-band (with
possible extension to H-band). We therefore avoid any unnecessary
splitting of light with dichroics. The narrow wavelength range also
improves the performance of multi-layer coatings and dichroics
typically by a factor of two. GRAVITY will use fluoride glass fibers
optimized for K-band operation. The attenuation of these fibers
(3~dB/km) is negligible for our fiber-length of a few meters, and the
fibers have already been successfully used in other interferometric
projects (e.g. 'OHANA, Perrin et al. 2004). GRAVITY will only provide
low and medium spectral resolution, because we can use high-throughput
transmissive dispersion elements for that. It is important to note
that GRAVITY does not intend to extend its beam combination to more
than 4 telescopes (4+4 strategy for VLTI), because any additional
combination is accompanied by according light losses. In order to
avoid extra background, a cold stop will be installed in the
fiber-coupler unit.  In addition, the beam-combiner and the subsequent
camera / spectrometer will be operated at cryogenic
temperatures. Several options are already available for the science
detector (e.g. HAWAII from Rockwell Inc., Finger et
al. 2002). However, new developments for IR AO wavefront-sensors
(e.g. Rockwell CALICO, Finger et al. 2004) will provide larger pixels
at even lower read-noise. With such detectors only one detector pixel
is necessary per beam-combiner output and spectral channel, reducing
the effective read-noise per information element accordingly. In
addition, our restriction to K-band will allow to optimize and select
detectors for best quantum efficiency at 2 $\mu$m. GRAVITY will
enclose all major components in an evacuated cryostat for optimum
cleanliness and stability. Also GRAVITY will be free of temperature-
and pressure variations, which minimizes the need for calibration, and
thus significantly increases the observing efficiency.

\subsection{Beam Combiner Options}
\label{beam combiner}
Three beam-combiner options are considered for GRAVITY: classical
bulk-optic combination, integrated-optics and fiber X-couplers. The
baseline is integrated optics. Widely used in the telecommunication
NIR bands (e.g. for beam switching), the technique has already been
applied in astronomical interferometry (IONIC, Berger et
al. 2003). Also the extension to the K-band has been demonstrated
(Laurent et al. 2002). A comparison of various concepts for the
combination of four beams can be found in Le Bouquin et al. (2004).
Beam combination in a bulk combiner can be designed to be very
efficient, and has successfully been used in astronomical
interferometry (e.g. COAST, Haniff et al. 2004). Dust contamination
and thermal drifts that might degrade the performance would be avoided
in the cryostat. The third option would be fiber X-couplers. The usage
of fluoride-glass fiber combiners for the K-band has been demonstrated
(e.g. FLUOR, Coud\'e du Foresto et al. 1998; VINCI, Kervella
et. al. 2000).


\section{Conclusions}
\label{conclusions}
GRAVITY will open a new window for unique science with the VLTI. In
addition to a wide variety of astronomical key-observations, the
instrument will --- for the first time --- directly probe the regime
of strong gravity close to black holes. The key-components of GRAVITY
are an IR wavefront-sensor, a two-object beam-combiner, and
high-throughput optics. As such GRAVITY will complement the
multi-mode, multi-wavelength, high spectral-resolution instruments
with simultaneous multi-baseline astrometry, optimum sensitivity for
faint objects, and a 10 times better astrometric accuracy for the
UTs. Thanks to its internal phase-referencing, GRAVITY will be able to
start operation already with the present PRIMA facility. 


\input{referenc}


\printindex
\end{document}

%% file: referenc.tex
%
%

%
%

%% file: gravity_220805.bbl
\begin{thebibliography}{99.}
%
%
%
\bibitem{arsenault2003} R. Arsenault et al: Proc. SPIE, \textbf{4839}, 174 (2003)
\bibitem{berger2003} J.-P. Berger et al.: Proc. SPIE, \textbf{4838}, 1099 (2003)
\bibitem{barvainis1987} R. Barvainis: ApJ, \textbf{320}, 537 (1987)
\bibitem{baumgardt2005} H. Baumgardt et al.: ApJ, \textbf{620}, 238 (2005)
\bibitem{bender2005} R. Bender et al.: ApJ, accepted (2005)
\bibitem{broderick2005}  A.E. Broderick \& A. Loeb: MNRAS, submitted, astro-ph/0506433 (2005)
\bibitem{chauvin2005} G. Chauvin et al.: A\&A, \textbf{438}, 25 (2005)
\bibitem{close2005} L.M. Close et al.: Nature, \textbf{433}, 286 (2005)
\bibitem{coude1998} V. Coud\'e du Foresto et al.: Proc. SPIE,
\textbf{3350}, 856 (1998)
\bibitem{cugat2001} O. Cugat et al.: Sensors and Actuators, \textbf{A 89}, 1 (2001)
\bibitem{dayton2000} S. Dayton et al.: Optics Communications, 339, 345 (2000)
\bibitem{dayton2001} S. Dayton et al.: OSA Optics Express, \textbf{8
No. 1}, 17 (2001)
\bibitem{delplancke2002} F. Delplancke et al.: Proc. SPIE, \textbf{4006}, 41 (2000)
\bibitem{delplancke2004a} F. Delplancke et al.: Reference missions for PRIMA, 
  Report prepared by the VLTI Implementation Committee (2004)
\bibitem{delplancke2004b} F. Delplancke: PRIMA Executive Summary,
  private communication (2004) 
\bibitem{eisenhauer2005} F. Eisenhauer et al.: ApJ, in press (2005)
\bibitem{elvis2002} M. Elvis \& M. Karovska: ApJ, \textbf{581}, L67 (2002)
\bibitem{finger2002} G. Finger et al.: Proc. SPIE, \textbf{4841}, 89 (2002)
\bibitem{finger2004} G. Finger et al.: Proc. SPIE, \textbf{5499}, 97 (2004)
\bibitem{gendron2003} E. Gendron et al.: Proc. SPIE, \textbf{4839},
  195 (2003)
\bibitem{genzel2003} R. Genzel et al.: Nature, \textbf{425}, 934 (2003)
\bibitem{glindemann2004} A. Glindemann et al.: Proc. SPIE, \textbf{5491},
  447 (2004) 
\bibitem{haniff2004} C.A. Haniff et al.: Proc. SPIE, \textbf{5491},
  59 (2004)
\bibitem{jaffe2004} W. Jaffe et al.: Nature, \textbf{429}, 47 (2004)
\bibitem{kaspi2000} S. Kaspi et al.: ApJ, \textbf{533}, 631 (2000)
\bibitem{kervella2000} P. Kervella et al.: Proc SPIE, \textbf{4006},
  31 (2000)
\bibitem{koehler2002} B. Koehler \& P. Gitton: Interface Control
  Document between VLTI and its Instruments, VLT-ICD-ESO-15000-1826,
  Issue 3.0 (2002)
\bibitem{lane2004} B.F. Lane \& M.W. Muterspaugh: ApJ, \textbf{601},
1129 (2004)
\bibitem{laurent2002} E. Laurent et al.: A\&A \textbf{390}, 1171 (2002)
\bibitem{leveque} S. Leveque et al.: Proc. SPIE, \textbf{4838}, 983 (2003)
\bibitem{lebouquin2004} J.B. Le Bouquin et al.: Proc. SPIE, \textbf{5491}, 1362 (2004)
\bibitem{maillard2004} J.P. Maillard et al.: A\&A, \textbf{423}, 155 (2004)
\bibitem{malbet1999} F. Malbet A\&AS , \textbf{138}, 135 (1999)
\bibitem{neuhaeuser2005} R. Neuh\"auser et al.: A\&A, \textbf{435}, 13 (2005)
\bibitem{ott1999} T. Ott et al.: ApJ, \textbf{523}, 248 (1999)
\bibitem{paumard2005a} T. Paumard et al.: this proceedings (2005a)
\bibitem{paumard2005b} T. Paumard et al.: A\&A, in preparation (2005b)
\bibitem{perrin2004} G. Perrin et al.: Proc. SPIE, \textbf{5491}, 391 (2004)
\bibitem{petrov2000} R. Petrov et al.: Proc. SPIE, \textbf{4006}, 68 (2000)
\bibitem{rantakyroe2005} F. Rantakyr\"o: AMBER user manual,
  VLT-MAN-ESO-15830-3522, Issue 1.2 (2005)
\bibitem{turner1999} N.J. Turner, et al.: ApJ, \textbf{524}, 129 (1999)
\bibitem{wilhelm2002} R. Wilhelm et al.: Functional Description of the
  VLTI, VLT-ICD-ESO-15000-1918, Issue 2.0 (2002)
\bibitem{woillez2005} J. Woillez et al.: in preperation (2005)
\end{thebibliography}
